\documentclass[prc,aps,twocolumn,floatfix,unsortedaddress,reprint]{revtex4-1}

\usepackage{graphics}
\usepackage{epsfig}
\usepackage{amsmath}
\usepackage{slashbox}
\usepackage{graphicx}

\usepackage{lineno}

\def\inbar{\,\vrule height1.5ex width.4pt depth0pt}
\def\IR{\relax{\rm I\kern-.18em R}}
\def\IC{\relax\hbox{$\inbar\kern-.3em{\rm C}$}}

\newcommand{\be}{\begin{linenomath}\begin{equation}}
\newcommand{\ee}{\end{equation}\end{linenomath}}
\newcommand{\beqa}{\begin{linenomath}\begin{eqnarray}}
\newcommand{\eeqa}{\end{eqnarray}\end{linenomath}}

\newcommand{\footfrac}[2]%

\newcommand{\open}{{<\kern -0.3em{\scriptscriptstyle )}}}

\begin{document}


\title{Updating spin dependent Regge intercepts}

\author{Steven D. Bass}
\affiliation{\mbox{Kitzb\"uhel Centre for Physics,
Kitzb\"uhel, Austria}}
\affiliation{\mbox{Marian Smoluchowski Institute of Physics, Jagiellonian University, 
PL 30-348 Krakow, Poland}}

\author{Magdalena Skurzok}
\author{Pawel Moskal}
\affiliation{\mbox{Marian Smoluchowski Institute of Physics, Jagiellonian University, 
PL 30-348 Krakow, Poland}}

\begin{abstract}
We use new high statistics data from CLAS and COMPASS on 
the nucleon's spin structure function at low Bjorken $x$ 
and low virtuality, $Q^2 < 0.5$ GeV$^2$,
together with earlier measurements from the SLAC E-143, 
HERMES and GDH experiments
to estimate the effective intercept(s) for spin dependent
Regge theory.
We find $\alpha_{a_1} = 0.31 \pm 0.04$ for the intercept describing the high-energy behaviour of spin dependent photoabsorption together with a new estimate for the 
high-energy part of the Gerasimov-Drell-Hearn sum-rule, 
$-15 \pm 2 \mu$b from photon-proton 
centre-of-mass energy greater than 2.5 GeV.
Our value of $\alpha_{a_1}$ suggests QCD physics beyond a 
simple straight-line $a_1$ trajectory.
\end{abstract}

\maketitle

\section{INTRODUCTION}

The high-energy behaviour of the spin dependent part of 
the photon-proton total cross section is important for 
determining the Gerasimov-Drell-Hearn sum-rule for polarised 
photoabsorption with real photons
\cite{Gerasimov:1965et,Drell:1966jv},
as well as studies of the transition 
from polarised photoproduction to deep inelastic scattering
\cite{Bass:2000zv}.

Here we investigate this behaviour using the new 
high statistics measurements from 
CLAS at Jefferson Laboratory~\cite{Fersch:2017qrq}
and COMPASS at CERN~\cite{Aghasyan:2017vck}
of the spin asymmetry for polarised photon-proton 
collisions at low photon virtuality $Q^2 < 0.5$ GeV$^2$ 
and centre-of-mass energy $\sqrt{s} \geq 2.5$ GeV,
together with
earlier measurements from 
the E-143 experiment at SLAC~\cite{Abe:1998wq}, 
HERMES at DESY~\cite{Airapetian:2006vy}
and the GDH Collaboration in Bonn~\cite{Dutz:2004zz}.

The large $s$ dependence of hadronic total cross-sections
is usually described in terms of 
Regge exchanges \cite{Collins:1984tj,Kuti:1997hz}, 
{\it e.g.} 
summing the exchanges of hadrons with given quantum numbers 
that occur along Regge trajectories with slope 
(often taken as a straight line) related to the confinement potential.
Regge phenomenology has considerable success in
describing unpolarised high-energy scattering processes
\cite{Landshoff:1994up}.

\section{SPIN DEPENDENT REGGE THEORY}

Let $\sigma_A$ and $\sigma_P$ denote the two cross-sections 
for the absorption of a transversely polarised photon with
spin antiparallel $\sigma_A$ or parallel $\sigma_P$ to the 
spin of the target nucleon.
The Regge prediction for the isovector and isoscalar parts 
of $(\sigma_A - \sigma_P)$ 
for a real photon, $Q^2=0$, 
with $s \rightarrow \infty$ 
is \cite{Heimann:1973hq,Bass:1994xb,Close:1994he}:
\begin{eqnarray}
& & \biggl( \sigma_A - \sigma_P \biggr)^{(p-n)}  
\sim \sum_i N^{(3)}_i s^{\alpha_{a_i -1}} 
\nonumber \\
& & 
\biggl( \sigma_A - \sigma_P \biggr)^{(p+n)} 
\sim
  \sum_i N^{(0)}_i  s^{\alpha_{f_i} - 1}
+ N_g {\ln {s / \mu^2} \over s}.
\end{eqnarray}
Here, the $\alpha_i$ denote the Regge intercepts for 
isovector $a_1$(1260) Regge exchange and the 
$a_1$-pomeron cuts \cite{Heimann:1973hq}.
The $\alpha_{f_i}$ denote the intercepts for the isoscalar 
$f_1(1285)$ and $f_1(1420)$ Regge trajectories and their
$f_1$-pomeron cuts.
The logarithm $\ln s/s$ term comes from two 
non-perturbative gluon exchange in the $t$-channel 
\cite{Bass:1994xb}
with a vector short-range exchange-potential \cite{Close:1994he}
and the mass parameter $\mu$ is taken as a typical hadronic scale. 
The coefficients $N^{(3)}_i$, $N^{(0)}_i$ and $N_g$ are to be determined from experiment.

If one makes the usual assumption that the $a_1$ 
Regge trajectories are straight lines parallel 
to the $(\rho, \omega)$ trajectories then one finds
$\alpha_{a_1} \simeq -0.4$ for the leading trajectory,
within the range of possible $\alpha_{a_1}$ 
values between -0.5 and zero
discussed in Ref.~\cite{Ellis:1988mn}.
Fitting straight line trajectories through the 
$a_1(1260)$ and $a_3(2030)$ states,
the $a_1(1640)$ and $a_3(2310)$ states, 
and the $f_1(1285)$ and $f_3(2050)$ states 
yields near parallel trajectories with slopes
0.79 GeV$^{-2}$, 0.76 GeV$^{-2}$ and 0.78 GeV$^{-2}$
respectively.
The two leading trajectories then have slightly lower intercepts,
$\alpha_{a_1} = -0.25$ and $\alpha_{f_1} = - 0.29$.
With this value of $\alpha_{a_1}$ the effective intercepts corresponding to the $a_1$ soft-pomeron cut and the 
$a_1$ hard-pomeron cut are -0.17 and +0.15 respectively
{\it if} one takes the soft pomeron with intercept 
1.0808 and hard pomeron proposed in 
Ref.~\cite{Donnachie:1998gm} 
with intercept 1.4 as two distinct exchanges.
Values of $\alpha_{a_1}$ close to zero could be achieved 
with curved Regge trajectories; 
$\alpha_{a_1} = -0.03 \pm 0.07$
is found in the model of Ref.~\cite{Brisudova:1999ut}.
For this value the intercepts of the $a_1$ soft-pomeron 
cut and the $a_1$ hard-pomeron cut are $\sim$ +0.05 and 
$\sim$ +0.37.

Before presenting our new results, we first recall the 
challenge of understanding the proton's internal spin 
structure in high $Q^2$ deep inelastic scattering and
$Q^2$ dependence of the intercepts 
$\alpha_i$ describing the asymptotic high-energy behaviour.

In deep inelastic kinematics the nucleon's $g_1$ spin structure 
function is related to $(\sigma_A - \sigma_P)$ by
\begin{equation}
\biggl(\sigma_A - \sigma_P \biggr) 
\simeq {4 \pi^2 \alpha_{\rm QED} \over p.q} g_1
\end{equation}
where $p$ and $q$ are the proton and photon four-momenta respectively and 
$\alpha_{\rm QED}$ is the electromagnetic coupling.
The Regge prediction for the isovector
$g_1^{p-n} = g_1^p - g_1^n$ 
at small Bjorken $x$ $(= Q^2/2.p.q)$ is
\begin{equation}
g_1^{p-n} 
\sim \sum_i N^{(3)}_i \ \biggl( \frac{1}{x} \biggr)^{\alpha_i}
\end{equation}
with all data taken at the same $Q^2$.
Eq.~(3) follows from
$s = (p+q)^2 = Q^2 \frac{(1-x)}{x} + M^2$
where $M$ is the proton mass 
and
$s \simeq Q^2 / x$ in the small $x$ limit.
There is 
possible $Q^2$ dependence in the $\alpha_i$ and $N^{(3)}_i$.
The COMPASS experiment found 
\begin{equation}
g_1^{p-n} \sim x^{-0.22 \pm 0.07}
\end{equation}
corresponding to an effective intercept
$\alpha_{a_1} (Q^2) = 0.22 \pm 0.07$ at $Q^2 = 3$ GeV$^2$,
with small $x$ data down to $x_{\rm min} \sim 0.004$
\cite{Alekseev:2010hc}.

The isoscalar spin structure function
$g_1^{p+n} \sim 0$ for $x < 0.03$ at deep inelastic $Q^2$
\cite{Aidala:2012mv},
in sharp contrast to the unpolarised structure function $F_2$
where the isosinglet part dominates through gluonic exchanges.
The proton spin puzzle, why the quark spin content of 
the proton is so small $\sim 0.3$, concerns the 
collapse of the isoscalar spin sum structure function 
to near zero at this small $x$.
The spin puzzle is now understood in terms of pion cloud effects 
with transfer of quark spin to orbital angular momentum in 
the pion cloud \cite{Bass:2009ed}, 
a modest polarised gluon correction 
$- 3 \frac{\alpha_s}{2 \pi} \Delta g$ 
with $\Delta g$ less than about 0.5 at the scale of the experiments \cite{Aidala:2012mv}, 
and a possible topological effect at $x=0$ \cite{Bass:2004xa}.

The observed rise in $g_1^{p-n}$ at deep inelastic values 
of $Q^2$ is required to reproduce the area under 
the fundamental Bjorken sum rule,
\begin{equation}
\int_0^1 dx g_1^{(p-n)} (x,Q^2) 
= \frac{g_A^{(3)}}{6} C_{\rm NS} (Q^2) .
\end{equation}
Here $g_A^{(3)} = 1.270 \pm 0.003$  
is the isovector axial-charge measured in neutron 
beta-decays and
$C_{NS}(Q^2)$
is the perturbative QCD Wilson coefficient, $\simeq 0.85$ 
with QCD coupling $\alpha_s = 0.3$~\cite{Aidala:2012mv}.
The Bjorken sum-rule is connected to pion physics and
chiral symmetry through the Goldberger-Treiman relation
$2 M g_A^{(3)} = f_{\pi} g_{\pi NN}$
where $f_{\pi}$ is the pion decay constant
and
$g_{\pi NN}$ is the pion-nucleon coupling constant.
The sum-rule has been confirmed in polarised deep inelastic scattering experiments at the level of 5\% 
\cite{Alekseev:2010hc}.
About 50\% of the sum-rule comes from $x$ values less than about 0.15.
The $g_1^{p-n}$ data is consistent with quark model and 
perturbative QCD predictions in the valence region $x > 0.2$ \cite{Bass:1999uj}.
The size of $g_A^{(3)}$ forces us to accept a large contribution from small $x$ and the observed rise 
in $g_1^{p - n}$ is required to fulfill this non-perturbative constraint.

Perturbative QCD evolution acts to push the weight of the
distribution to smaller Bjorken $x$ with increasing $Q^2$
with perturbative  
calculations predicting rising $g_1^{p-n}$ at small $x$
and deep inelastic $Q^2$
\cite{Blumlein:1995jp,Kwiecinski:1999sk}.
Regge phenomenology should describe the high-energy part of 
$g_1$ close to photoproduction and provide the input for 
perturbative QCD evolution at deep inelastic values of $Q^2$.
One then applies perturbative QCD,
typically above $Q^2 > 1$ GeV$^2$.
These perturbative QCD calculations 
involve DGLAP evolution and double logarithm, 
$\alpha_s^m \ln^n \frac{1}{x}$, resummation at small $x$
\cite{Blumlein:2012bf},
in possible combination with vector meson dominance terms 
at low $Q^2$ \cite{Badelek:2002jr}.
For $g_1^{p-n}$ with DGLAP evolution
this approach has the challenging feature that 
the input and output (at soft and hard scales) 
are governed by non-perturbative constraints 
with perturbative QCD evolution in the middle
unless the $a_1$ Regge input has information 
about $g_A^{(3)}$ and chiral symmetry built into it.
One possibility is a separate hard-exchange 
contribution (perhaps an $a_1$ hard-pomeron cut) 
in addition to the soft $a_1$ term \cite{Bass:2006dq}.

\section{FITTING THE HIGH ENERGY SPIN ASYMMETRY}

We next estimate the spin dependent Regge intercepts.
Good statistics measurements of the
spin asymmetry for photon-proton collisions
$A_1^p = (\sigma_A - \sigma_P) / (\sigma_A + \sigma_P)$ 
at large $\sqrt{s}$ and low $Q^2$ have recently become available from the CLAS and COMPASS experiments, 
complementing earlier measurements from 
SLAC, HERMES and the GDH Collaboration.
We make a Regge motivated fit to this data on 
$
\Delta \sigma 
= \sigma_A - \sigma_P = A_1^p \ (\sigma_A + \sigma_P)
$
with the constraints
$\sqrt{s} \geq 2.5$ GeV
where Regge theory is expected 
to apply~\cite{Landshoff:1994up} 
and $Q^2 < 0.5$ GeV$^2$.
Keeping $Q^2 < 0.5$ GeV$^2$ is a compromise between keeping $Q^2$ as low as possible and including the maximum amount of data.
This input data involves 
18 points from COMPASS 
with $\sqrt{s}$ between 11 and 15 GeV \cite{Aghasyan:2017vck},
2 data points from HERMES with $\sqrt{s}$ at 6.6 and 6.8 GeV 
\cite{Airapetian:2006vy},
7 points from SLAC E-143 with $\sqrt{s}$ between 2.5 and 3.1 GeV \cite{Abe:1998wq},
and
102 points from CLAS between 2.5 and 2.9 GeV 
\cite{Fersch:2017qrq}.
This data is consistent with $A_1^p$ being $Q^2$ independent in each experiment within the chosen kinematics.
We also consider the highest energy single data point 
from the GDH photoproduction experiment with $\sqrt{s} = 2.5$ GeV and $Q^2=0$ \cite{Dutz:2004zz}.
Data at higher $Q^2$ values between 0.5 and 1 GeV$^2$
are in principle sensitive to the extra effects 
of turning on DGLAP evolution 
and decay of higher-twist terms with increasing $Q^2$.

The unpolarised total cross-section,
$\sigma_{\rm tot} = \sigma_A + \sigma_P$, 
measurements from HERA were found to be well described
by a combined Regge and 
Generalized Vector Meson Dominance (GVMD) motivated fit
in the kinematics $Q^2 < 0.65$ GeV$^2$ and $s \geq 3$GeV$^2$ 
\cite{Breitweg:2000yn,Breitweg:1998dz,Adloff:1997mf}.
The ZEUS Collaboration used the 4 parameter fit 
\cite{Breitweg:2000yn}
\begin{equation}
\sigma_{\rm tot}^{\gamma^* p} (s, Q^2) =
\biggl( {M_0^2 \over M_0^2 + Q^2} \biggr)
\biggl( A_R s^{\alpha_R-1} + A_P s^{\alpha_P-1} \biggr)
\end{equation}
to describe the low $Q^2$ region, 
also including fixed target data 
from the E665 Collaboration~\cite{Adams:1996gu},
with
$A_R = 147.8 \pm 4.6 \mu$b, $\alpha_R=0.5$ (fixed),
$A_P =  62.0 \pm 2.3 \mu$b,
$\alpha_P = 1.102 \pm 0.007$ and 
$M_0^2 = 0.52 \pm 0.04$GeV$^2$.

In the HERA kinematical region the total $\gamma^* p$ 
cross-section is related to $F_2(x,Q^2)$ by
\begin{equation}
\sigma_{\rm tot}^{\gamma^* p} (s, Q^2) 
\simeq {4 \pi^2 \alpha_{\rm QED} \over Q^2} F_2 (x,Q^2) 
\end{equation}
where $s \simeq Q^2/x$.
For $Q^2$ larger than 1 GeV$^2$ the HERA data on 
$F_2$ seems to be well described by DGLAP evolution.
Parametrising $F_2 \sim A x^{- \lambda}$ at small $x$ 
the effective intercept $\lambda$ is observed to grow
from $0.11 \pm 0.02$ at $Q^2=0.3$GeV$^2$ 
to $0.18 \pm 0.03$ at $Q^2=3.5$GeV$^2$, $0.31 \pm 0.02$ at 
35 GeV$^2$ and increases with increasing $Q^2$ 
\cite{Breitweg:1998dz,Adloff:1997mf,Desgrolard:1999ax}.
The value 0.4 was found at the highest $Q^2$ motivating
suggestions of a new hard pomeron 
\cite{Donnachie:1998gm,Donnachie:2001zt}.

Here, 
we first assume $A_1^p$ to be $Q^2$ independent in our chosen kinematics with 
$Q^2 < 0.5$GeV$^2$.
That is, we conjecture
\begin{equation}
(\sigma_A - \sigma_P)^{\gamma^* p}(s,Q^2) = 
\biggl( {M_0^2 \over M_0^2 + Q^2} \biggr) 
\ 
(\sigma_A - \sigma_P)^{\gamma p}(s,0)
\end{equation}
at large $s$ and small $Q^2$
with the same value of $M_0^2$ in both Eqs.(6) and (8)
and $Q^2$ independent values of the spin Regge intercepts
$\alpha_i$ at this low $Q^2$.

Second, we assume that the isoscalar deuteron asymmetry
$A_1^d$ can be taken as zero in first approximation.
The deuteron data on $A_1^d$ are consistent with zero 
in each experiment in our chosen kinematics
\cite{Abe:1998wq,Guler:2015hsw,Airapetian:2006vy,Ageev:2007du}
(as well as in $g_1^d$ measurements at deep inelastic 
 $Q^2$ and low $x < 0.03$ \cite{Aidala:2012mv}).
This means that we set the normalisation factors
$N^{(0)}_i = N_g = 0$ in Eq.(1).

Third, we take $\sigma_{\rm tot}$ from a fit to unpolarised data. 
We assume that the errors on $\sigma_{\rm tot}$ can be neglected compared to the errors on $A_1^p$.
For the total photoproduction cross-section we take
\begin{equation}
(\sigma_A + \sigma_P) 
= 67.7 \ s^{+0.0808} + 129 \ s^{-0.4545}
\label{sigtot}
\end{equation}
(in units of $\mu$b),
which provides a good Regge fit 
for $\sqrt{s}$ between 2.5 GeV and 250 GeV
\cite{Landshoff:1994up}.
The $s^{+0.0808}$ contribution is associated with 
 gluonic pomeron exchange and the $s^{-0.4545}$ contribution 
 is associated with the isoscalar $\omega$ and isovector 
 $\rho$ trajectories.

\begin{figure}[!t]  
\centerline{\includegraphics[width=0.50\textwidth]{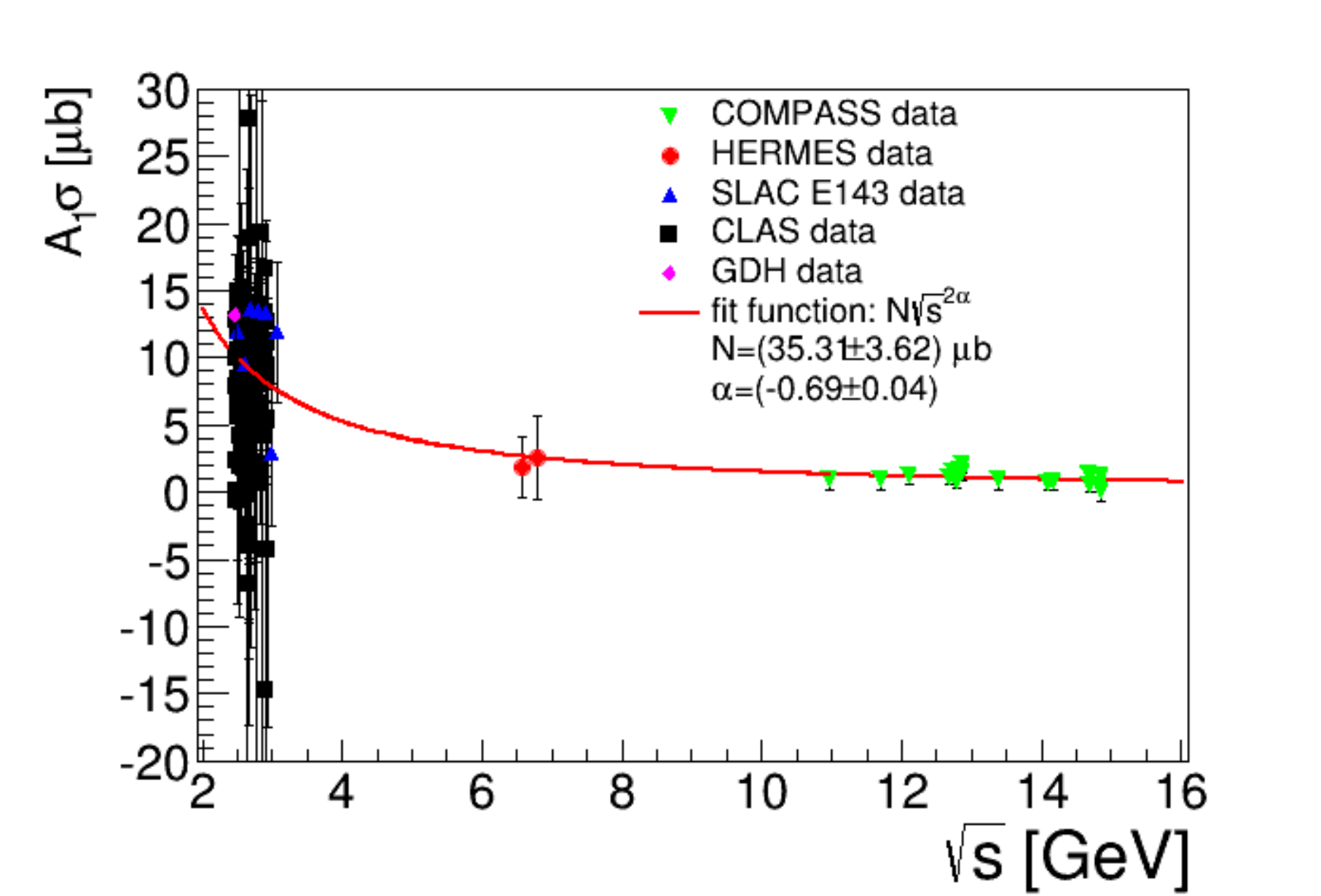}}
\caption{ 
Regge fit to 
$(\sigma_A - \sigma_P) = A_1^p \ (\sigma_A + \sigma_P)$
with spin data from the 
CLAS~\cite{Fersch:2017qrq},
COMPASS~\cite{Aghasyan:2017vck},
GDH~\cite{Dutz:2004zz},
HERMES~\cite{Airapetian:2006vy},
and SLAC E-143~\cite{Abe:1998wq}
experiments with $Q^2 < 0.5$GeV$^2$.
}
\end{figure}

Our best fit of form 
$(\sigma_A - \sigma_P) \sim N s^{\alpha}$
including all data is
\begin{equation}
(\sigma_A - \sigma_P)
= (35.3 \pm 3.6) \ s^{-0.69 \pm 0.04} \ \mu {\rm b}
\end{equation}
for $\sqrt{s} \geq 2.5$ GeV
corresponding to an effective Regge intercept 
\begin{equation}
\alpha_{a_1} = + 0.31 \pm 0.04
\end{equation}
-- see Fig.~1.
The $\chi^2/{\rm ndf}$ for the fit is 0.98.
Statistical and systematic errors for each data point
have been added in quadrature.

To convert the fit results in Eqs.~(9-11) into a prediction
for the asymmetry $A_1^p$ as a function of $x$,
it is important to note that $s \simeq Q^2/x$ 
at large centre-of-mass energy and 
take
into account that experimental measurements in different
$x$ bins are typically taken at different $Q^2$ values.
For example, the COMPASS measurements 
using a 160 GeV muon beam at
$\langle x \rangle = 0.000052$ were taken at 
$\langle Q^2 \rangle = 0.0062$ GeV$^2$
whereas
their measurements at 
$\langle x \rangle = 0.0020$ were taken at 
$\langle Q^2 \rangle = 0.33$ GeV$^2$~\cite{Aghasyan:2017vck},
a factor of 53 greater in $Q^2$.
Within each $x$ bin taken separately, $Q^2$ 
was varied over a more limited factor of 
about 5 and the experimental uncertainties 
too large to make definite conclusions about 
possible $Q^2$ dependence within individual $x$ bins.
All of our COMPASS points with $Q^2 < 0.5$ GeV$^2$ 
are in the range $\sqrt{s}$ between 11 and 15 GeV. 
One expects $A_1^p$ to vanish in the small $x$ limit, 
which follows in this 
data when all
points are shifted to the same $Q^2$ by dividing out
the factor $(Q^2)^{\alpha_{a_1}-1.0808} \sim (Q^2)^{-0.77}$
from Eqs.~(9-11).

\section{DISCUSSION}

It is very interesting that the intercept in Eq.(11) 
is close to the value found in deep inelastic scattering,
{\it viz.} $\alpha_{a_1} (Q^2) = 0.22 \pm 0.07$ 
at $Q^2 = 3$GeV$^2$ in Eq.(4).
Our new low $Q^2$ value signifies either the presence of 
a hard exchange, perhaps involving an $a_1$ hard pomeron 
cut,
or a curved Regge trajectory instead of just 
a simple straight-line $a_1$ Regge trajectory.

More valuable experimental input could come from 
the proposed future electron-ion-collider 
which could extend
the experimental data up to $\sqrt{s}$ 
values between 40 GeV and 140 GeV
\cite{Deshpande:2005wd,Deshpande:2017edt}
-- that is,
up to an order of magnitude higher in $\sqrt{s}$
than the 
present highest centre-of-mass energy COMPASS data.
Estimates for the expected asymmetries 
are given in \cite{Bass:2000zv}.
The fit values in Eqs.~(10, 11) suggest low $Q^2$ 
asymmetries 
$A_1^p = (1.7 \pm 0.5) \times 10^{-3}$ at $\sqrt{s} = 40$ GeV
and
$A_1^p = (2.5 \pm 1.0) \times 10^{-4}$ at $\sqrt{s} = 140$ GeV.

Taking the fit values in Eq.~(10), we estimate the 
high-energy contribution to the 
Gerasimov-Drell-Hearn sum-rule from 
$\sqrt{s} \geq 2.5$ GeV to be 
\begin{equation}
\int^{\infty}_{s_0} \frac{ds}{s - M^2} 
(\sigma_P - \sigma_A) 
= - 15 \pm 2 \ \mu {\rm b} .
\end{equation}
This determination compares with previous estimates:
$- 15 \pm 10 \mu$b 
for $\sqrt{s} \geq 2.5$ GeV
based on an extrapolation of lower energy photoproduction 
data
which also gave 
$\alpha_{a_1} = 0.42 \pm 0.23$~\cite{Helbing:2006zp},
$- 25 \pm 10 \mu$b 
from an early estimate using lower statistics low $Q^2$ 
data (pre CLAS and COMPASS) 
for $\sqrt{s} \geq 2.5$GeV \cite{Bass:1997fh}, 
and $- 26 \pm 7 \mu$b 
for $\sqrt{s} \geq 2$ GeV \cite{Bianchi:1999qs}
from early Regge fits to low $Q^2$ data. 
The new result in Eq.~(12) is a factor of 3.5 times 
more accurate than the previous most accurate determination.
The corresponding integral from threshold up to 
$\sqrt{s} = 2.5$ GeV 
has been extracted from proton fixed target 
experiments with photon energy up to 2.9 GeV.
One finds
$226 \pm 5 \pm 12 \mu$b \cite{Helbing:2006zp,Dutz:2004zz}.
Combining this number and the result in Eq.~(12)
gives 
\begin{equation}
\int^{\infty}_{M^2} 
\frac{ds}{s - M^2}
(\sigma_P - \sigma_A)
= 211 \pm 13 \ \mu {\rm b} 
\end{equation}
for the Gerasimov-Drell-Hearn sum-rule.
This value compares with the sum-rule prediction
$2 \pi^2 \alpha_{\rm QED} \kappa^2 / M^2
 = 205 \ \mu$b
with $\kappa=1.79$ the proton's anomalous magnetic moment
\cite{Gerasimov:1965et,Drell:1966jv}.

\section*{Acknowledgments}

We thank R. Fersch for helpful communications.

\bibliography{regge}

\begin{thebibliography}{40}%
\makeatletter
\providecommand \@ifxundefined [1]{%
 \@ifx{#1\undefined}
}%
\providecommand \@ifnum [1]{%
 \ifnum #1\expandafter \@firstoftwo
 \else \expandafter \@secondoftwo
 \fi
}%
\providecommand \@ifx [1]{%
 \ifx #1\expandafter \@firstoftwo
 \else \expandafter \@secondoftwo
 \fi
}%
\providecommand \natexlab [1]{#1}%
\providecommand \enquote  [1]{``#1''}%
\providecommand \bibnamefont  [1]{#1}%
\providecommand \bibfnamefont [1]{#1}%
\providecommand \citenamefont [1]{#1}%
\providecommand \href@noop [0]{\@secondoftwo}%
\providecommand \href [0]{\begingroup \@sanitize@url \@href}%
\providecommand \@href[1]{\@@startlink{#1}\@@href}%
\providecommand \@@href[1]{\endgroup#1\@@endlink}%
\providecommand \@sanitize@url [0]{\catcode `\\12\catcode `\$12\catcode
  `\&12\catcode `\#12\catcode `\^12\catcode `\_12\catcode `\%12\relax}%
\providecommand \@@startlink[1]{}%
\providecommand \@@endlink[0]{}%
\providecommand \url  [0]{\begingroup\@sanitize@url \@url }%
\providecommand \@url [1]{\endgroup\@href {#1}{\urlprefix }}%
\providecommand \urlprefix  [0]{URL }%
\providecommand \Eprint [0]{\href }%
\providecommand \doibase [0]{http://dx.doi.org/}%
\providecommand \selectlanguage [0]{\@gobble}%
\providecommand \bibinfo  [0]{\@secondoftwo}%
\providecommand \bibfield  [0]{\@secondoftwo}%
\providecommand \translation [1]{[#1]}%
\providecommand \BibitemOpen [0]{}%
\providecommand \bibitemStop [0]{}%
\providecommand \bibitemNoStop [0]{.\EOS\space}%
\providecommand \EOS [0]{\spacefactor3000\relax}%
\providecommand \BibitemShut  [1]{\csname bibitem#1\endcsname}%
\let\auto@bib@innerbib\@empty
\bibitem [{\citenamefont {Gerasimov}(1966)}]{Gerasimov:1965et}%
  \BibitemOpen
  \bibfield  {author} {\bibinfo {author} {\bibfnamefont {S.~B.}\ \bibnamefont
  {Gerasimov}},\ }\href@noop {} {\bibfield  {journal} {\bibinfo  {journal}
  {Sov. J. Nucl. Phys.}\ }\textbf {\bibinfo {volume} {2}},\ \bibinfo {pages}
  {430} (\bibinfo {year} {1966})},\ \bibinfo {note} {[Yad.
  Fiz.2,598(1965)]}\BibitemShut {NoStop}%
\bibitem [{\citenamefont {Drell}\ and\ \citenamefont
  {Hearn}(1966)}]{Drell:1966jv}%
  \BibitemOpen
  \bibfield  {author} {\bibinfo {author} {\bibfnamefont {S.~D.}\ \bibnamefont
  {Drell}}\ and\ \bibinfo {author} {\bibfnamefont {A.~C.}\ \bibnamefont
  {Hearn}},\ }\href {\doibase 10.1103/PhysRevLett.16.908} {\bibfield  {journal}
  {\bibinfo  {journal} {Phys. Rev. Lett.}\ }\textbf {\bibinfo {volume} {16}},\
  \bibinfo {pages} {908} (\bibinfo {year} {1966})}\BibitemShut {NoStop}%
\bibitem [{\citenamefont {Bass}\ and\ \citenamefont
  {De~Roeck}(2001)}]{Bass:2000zv}%
  \BibitemOpen
  \bibfield  {author} {\bibinfo {author} {\bibfnamefont {S.~D.}\ \bibnamefont
  {Bass}}\ and\ \bibinfo {author} {\bibfnamefont {A.}~\bibnamefont
  {De~Roeck}},\ }\href {\doibase 10.1007/s100520000547} {\bibfield  {journal}
  {\bibinfo  {journal} {Eur. Phys. J.}\ }\textbf {\bibinfo {volume} {C18}},\
  \bibinfo {pages} {531} (\bibinfo {year} {2001})}\BibitemShut {NoStop}%
\bibitem [{\citenamefont {Fersch}\ \emph {et~al.}(2017)\citenamefont {Fersch}
  \emph {et~al.}}]{Fersch:2017qrq}%
  \BibitemOpen
  \bibfield  {author} {\bibinfo {author} {\bibfnamefont {R.}~\bibnamefont
  {Fersch}} \emph {et~al.} (\bibinfo {collaboration} {CLAS}),\ }\href {\doibase
  10.1103/PhysRevC.96.065208} {\bibfield  {journal} {\bibinfo  {journal} {Phys.
  Rev.}\ }\textbf {\bibinfo {volume} {C96}},\ \bibinfo {pages} {065208}
  (\bibinfo {year} {2017})}\BibitemShut {NoStop}%
\bibitem [{\citenamefont {Aghasyan}\ \emph {et~al.}(2018)\citenamefont
  {Aghasyan} \emph {et~al.}}]{Aghasyan:2017vck}%
  \BibitemOpen
  \bibfield  {author} {\bibinfo {author} {\bibfnamefont {M.}~\bibnamefont
  {Aghasyan}} \emph {et~al.} (\bibinfo {collaboration} {COMPASS}),\ }\href
  {\doibase 10.1016/j.physletb.2018.03.044} {\bibfield  {journal} {\bibinfo
  {journal} {Phys. Lett.}\ }\textbf {\bibinfo {volume} {B781}},\ \bibinfo
  {pages} {464} (\bibinfo {year} {2018})}\BibitemShut {NoStop}%
\bibitem [{\citenamefont {Abe}\ \emph {et~al.}(1998)\citenamefont {Abe} \emph
  {et~al.}}]{Abe:1998wq}%
  \BibitemOpen
  \bibfield  {author} {\bibinfo {author} {\bibfnamefont {K.}~\bibnamefont
  {Abe}} \emph {et~al.} (\bibinfo {collaboration} {E143}),\ }\href {\doibase
  10.1103/PhysRevD.58.112003} {\bibfield  {journal} {\bibinfo  {journal} {Phys.
  Rev.}\ }\textbf {\bibinfo {volume} {D58}},\ \bibinfo {pages} {112003}
  (\bibinfo {year} {1998})},\ \bibinfo {note} {[Full data are at
  www.slac.stanford.edu/exp/e143/]}\BibitemShut {NoStop}%
\bibitem [{\citenamefont {Airapetian}\ \emph {et~al.}(2007)\citenamefont
  {Airapetian} \emph {et~al.}}]{Airapetian:2006vy}%
  \BibitemOpen
  \bibfield  {author} {\bibinfo {author} {\bibfnamefont {A.}~\bibnamefont
  {Airapetian}} \emph {et~al.} (\bibinfo {collaboration} {HERMES}),\ }\href
  {\doibase 10.1103/PhysRevD.75.012007} {\bibfield  {journal} {\bibinfo
  {journal} {Phys. Rev.}\ }\textbf {\bibinfo {volume} {D75}},\ \bibinfo {pages}
  {012007} (\bibinfo {year} {2007})}\BibitemShut {NoStop}%
\bibitem [{\citenamefont {Dutz}\ \emph {et~al.}(2004)\citenamefont {Dutz} \emph
  {et~al.}}]{Dutz:2004zz}%
  \BibitemOpen
  \bibfield  {author} {\bibinfo {author} {\bibfnamefont {H.}~\bibnamefont
  {Dutz}} \emph {et~al.},\ }\href {\doibase 10.1103/PhysRevLett.93.032003}
  {\bibfield  {journal} {\bibinfo  {journal} {Phys. Rev. Lett.}\ }\textbf
  {\bibinfo {volume} {93}},\ \bibinfo {pages} {032003} (\bibinfo {year}
  {2004})}\BibitemShut {NoStop}%
\bibitem [{\citenamefont {Collins}\ and\ \citenamefont
  {Martin}(1984)}]{Collins:1984tj}%
  \BibitemOpen
  \bibfield  {author} {\bibinfo {author} {\bibfnamefont {P.~D.~B.}\
  \bibnamefont {Collins}}\ and\ \bibinfo {author} {\bibfnamefont {A.~D.}\
  \bibnamefont {Martin}},\ }\href@noop {} {\emph {\bibinfo {title} {{Hadron
  Interactions}}}}\ (\bibinfo  {publisher} {Adam Hilger, Bristol U.~K.},\
  \bibinfo {year} {1984})\BibitemShut {NoStop}%
\bibitem [{\citenamefont {Kuti}(1997)}]{Kuti:1997hz}%
  \BibitemOpen
  \bibfield  {author} {\bibinfo {author} {\bibfnamefont {J.}~\bibnamefont
  {Kuti}},\ }\href@noop {} {\bibfield  {journal} {\bibinfo  {journal} {Acta
  Phys. Hung.}\ }\textbf {\bibinfo {volume} {A5}},\ \bibinfo {pages} {195}
  (\bibinfo {year} {1997})}\BibitemShut {NoStop}%
\bibitem [{\citenamefont {Landshoff}(1994)}]{Landshoff:1994up}%
  \BibitemOpen
  \bibfield  {author} {\bibinfo {author} {\bibfnamefont {P.~V.}\ \bibnamefont
  {Landshoff}},\ }in\ \href@noop {} {\emph {\bibinfo {booktitle} {{Proceedings:
  Summer School on Hadronic Aspects of Collider Physics, Zuoz, Switzerland, Aug
  23-31, 1994}}}}\ (\bibinfo {year} {1994})\ pp.\ \bibinfo {pages} {135--150},\
  \Eprint {http://arxiv.org/abs/hep-ph/9410250} {arXiv:hep-ph/9410250 [hep-ph]}
  \BibitemShut {NoStop}%
\bibitem [{\citenamefont {Heimann}(1973)}]{Heimann:1973hq}%
  \BibitemOpen
  \bibfield  {author} {\bibinfo {author} {\bibfnamefont {R.~L.}\ \bibnamefont
  {Heimann}},\ }\href {\doibase 10.1016/0550-3213(73)90635-4} {\bibfield
  {journal} {\bibinfo  {journal} {Nucl. Phys.}\ }\textbf {\bibinfo {volume}
  {B64}},\ \bibinfo {pages} {429} (\bibinfo {year} {1973})}\BibitemShut
  {NoStop}%
\bibitem [{\citenamefont {Bass}\ and\ \citenamefont
  {Landshoff}(1994)}]{Bass:1994xb}%
  \BibitemOpen
  \bibfield  {author} {\bibinfo {author} {\bibfnamefont {S.~D.}\ \bibnamefont
  {Bass}}\ and\ \bibinfo {author} {\bibfnamefont {P.~V.}\ \bibnamefont
  {Landshoff}},\ }\href {\doibase 10.1016/0370-2693(94)90569-X} {\bibfield
  {journal} {\bibinfo  {journal} {Phys. Lett.}\ }\textbf {\bibinfo {volume}
  {B336}},\ \bibinfo {pages} {537} (\bibinfo {year} {1994})}\BibitemShut
  {NoStop}%
\bibitem [{\citenamefont {Close}\ and\ \citenamefont
  {Roberts}(1994)}]{Close:1994he}%
  \BibitemOpen
  \bibfield  {author} {\bibinfo {author} {\bibfnamefont {F.~E.}\ \bibnamefont
  {Close}}\ and\ \bibinfo {author} {\bibfnamefont {R.~G.}\ \bibnamefont
  {Roberts}},\ }\href {\doibase 10.1016/0370-2693(94)90245-3} {\bibfield
  {journal} {\bibinfo  {journal} {Phys. Lett.}\ }\textbf {\bibinfo {volume}
  {B336}},\ \bibinfo {pages} {257} (\bibinfo {year} {1994})}\BibitemShut
  {NoStop}%
\bibitem [{\citenamefont {Ellis}\ and\ \citenamefont
  {Karliner}(1988)}]{Ellis:1988mn}%
  \BibitemOpen
  \bibfield  {author} {\bibinfo {author} {\bibfnamefont {J.~R.}\ \bibnamefont
  {Ellis}}\ and\ \bibinfo {author} {\bibfnamefont {M.}~\bibnamefont
  {Karliner}},\ }\href {\doibase 10.1016/0370-2693(88)91050-7} {\bibfield
  {journal} {\bibinfo  {journal} {Phys. Lett.}\ }\textbf {\bibinfo {volume}
  {B213}},\ \bibinfo {pages} {73} (\bibinfo {year} {1988})}\BibitemShut
  {NoStop}%
\bibitem [{\citenamefont {Donnachie}\ and\ \citenamefont
  {Landshoff}(1998)}]{Donnachie:1998gm}%
  \BibitemOpen
  \bibfield  {author} {\bibinfo {author} {\bibfnamefont {A.}~\bibnamefont
  {Donnachie}}\ and\ \bibinfo {author} {\bibfnamefont {P.~V.}\ \bibnamefont
  {Landshoff}},\ }\href {\doibase 10.1016/S0370-2693(98)00899-5} {\bibfield
  {journal} {\bibinfo  {journal} {Phys. Lett.}\ }\textbf {\bibinfo {volume}
  {B437}},\ \bibinfo {pages} {408} (\bibinfo {year} {1998})}\BibitemShut
  {NoStop}%
\bibitem [{\citenamefont {Brisudova}\ \emph {et~al.}(2000)\citenamefont
  {Brisudova}, \citenamefont {Burakovsky},\ and\ \citenamefont
  {Goldman}}]{Brisudova:1999ut}%
  \BibitemOpen
  \bibfield  {author} {\bibinfo {author} {\bibfnamefont {M.~M.}\ \bibnamefont
  {Brisudova}}, \bibinfo {author} {\bibfnamefont {L.}~\bibnamefont
  {Burakovsky}}, \ and\ \bibinfo {author} {\bibfnamefont {T.}~\bibnamefont
  {Goldman}},\ }\href {\doibase 10.1103/PhysRevD.61.054013} {\bibfield
  {journal} {\bibinfo  {journal} {Phys. Rev.}\ }\textbf {\bibinfo {volume}
  {D61}},\ \bibinfo {pages} {054013} (\bibinfo {year} {2000})}\BibitemShut
  {NoStop}%
\bibitem [{\citenamefont {Alekseev}\ \emph {et~al.}(2010)\citenamefont
  {Alekseev} \emph {et~al.}}]{Alekseev:2010hc}%
  \BibitemOpen
  \bibfield  {author} {\bibinfo {author} {\bibfnamefont {M.~G.}\ \bibnamefont
  {Alekseev}} \emph {et~al.} (\bibinfo {collaboration} {COMPASS}),\ }\href
  {\doibase 10.1016/j.physletb.2010.05.069} {\bibfield  {journal} {\bibinfo
  {journal} {Phys. Lett.}\ }\textbf {\bibinfo {volume} {B690}},\ \bibinfo
  {pages} {466} (\bibinfo {year} {2010})}\BibitemShut {NoStop}%
\bibitem [{\citenamefont {Aidala}\ \emph {et~al.}(2013)\citenamefont {Aidala},
  \citenamefont {Bass}, \citenamefont {Hasch},\ and\ \citenamefont
  {Mallot}}]{Aidala:2012mv}%
  \BibitemOpen
  \bibfield  {author} {\bibinfo {author} {\bibfnamefont {C.~A.}\ \bibnamefont
  {Aidala}}, \bibinfo {author} {\bibfnamefont {S.~D.}\ \bibnamefont {Bass}},
  \bibinfo {author} {\bibfnamefont {D.}~\bibnamefont {Hasch}}, \ and\ \bibinfo
  {author} {\bibfnamefont {G.~K.}\ \bibnamefont {Mallot}},\ }\href {\doibase
  10.1103/RevModPhys.85.655} {\bibfield  {journal} {\bibinfo  {journal} {Rev.
  Mod. Phys.}\ }\textbf {\bibinfo {volume} {85}},\ \bibinfo {pages} {655}
  (\bibinfo {year} {2013})}\BibitemShut {NoStop}%
\bibitem [{\citenamefont {Bass}\ and\ \citenamefont
  {Thomas}(2010)}]{Bass:2009ed}%
  \BibitemOpen
  \bibfield  {author} {\bibinfo {author} {\bibfnamefont {S.~D.}\ \bibnamefont
  {Bass}}\ and\ \bibinfo {author} {\bibfnamefont {A.~W.}\ \bibnamefont
  {Thomas}},\ }\href {\doibase 10.1016/j.physletb.2010.01.008} {\bibfield
  {journal} {\bibinfo  {journal} {Phys. Lett.}\ }\textbf {\bibinfo {volume}
  {B684}},\ \bibinfo {pages} {216} (\bibinfo {year} {2010})}\BibitemShut
  {NoStop}%
\bibitem [{\citenamefont {Bass}(2005)}]{Bass:2004xa}%
  \BibitemOpen
  \bibfield  {author} {\bibinfo {author} {\bibfnamefont {S.~D.}\ \bibnamefont
  {Bass}},\ }\href {\doibase 10.1103/RevModPhys.77.1257} {\bibfield  {journal}
  {\bibinfo  {journal} {Rev. Mod. Phys.}\ }\textbf {\bibinfo {volume} {77}},\
  \bibinfo {pages} {1257} (\bibinfo {year} {2005})}\BibitemShut {NoStop}%
\bibitem [{\citenamefont {Bass}(1999)}]{Bass:1999uj}%
  \BibitemOpen
  \bibfield  {author} {\bibinfo {author} {\bibfnamefont {S.~D.}\ \bibnamefont
  {Bass}},\ }\href {\doibase 10.1007/s100500050252} {\bibfield  {journal}
  {\bibinfo  {journal} {Eur. Phys. J.}\ }\textbf {\bibinfo {volume} {A5}},\
  \bibinfo {pages} {17} (\bibinfo {year} {1999})}\BibitemShut {NoStop}%
\bibitem [{\citenamefont {Bl{\"u}mlein}\ and\ \citenamefont
  {Vogt}(1996)}]{Blumlein:1995jp}%
  \BibitemOpen
  \bibfield  {author} {\bibinfo {author} {\bibfnamefont {J.}~\bibnamefont
  {Bl{\"u}mlein}}\ and\ \bibinfo {author} {\bibfnamefont {A.}~\bibnamefont
  {Vogt}},\ }\href {\doibase 10.1016/0370-2693(95)01568-X} {\bibfield
  {journal} {\bibinfo  {journal} {Phys. Lett.}\ }\textbf {\bibinfo {volume}
  {B370}},\ \bibinfo {pages} {149} (\bibinfo {year} {1996})}\BibitemShut
  {NoStop}%
\bibitem [{\citenamefont {Kwiecinski}\ and\ \citenamefont
  {Ziaja}(1999)}]{Kwiecinski:1999sk}%
  \BibitemOpen
  \bibfield  {author} {\bibinfo {author} {\bibfnamefont {J.}~\bibnamefont
  {Kwiecinski}}\ and\ \bibinfo {author} {\bibfnamefont {B.}~\bibnamefont
  {Ziaja}},\ }\href {\doibase 10.1103/PhysRevD.60.054004} {\bibfield  {journal}
  {\bibinfo  {journal} {Phys. Rev.}\ }\textbf {\bibinfo {volume} {D60}},\
  \bibinfo {pages} {054004} (\bibinfo {year} {1999})}\BibitemShut {NoStop}%
\bibitem [{\citenamefont {Bl{\"u}mlein}(2013)}]{Blumlein:2012bf}%
  \BibitemOpen
  \bibfield  {author} {\bibinfo {author} {\bibfnamefont {J.}~\bibnamefont
  {Bl{\"u}mlein}},\ }\href {\doibase 10.1016/j.ppnp.2012.09.006} {\bibfield
  {journal} {\bibinfo  {journal} {Prog. Part. Nucl. Phys.}\ }\textbf {\bibinfo
  {volume} {69}},\ \bibinfo {pages} {28} (\bibinfo {year} {2013})}\BibitemShut
  {NoStop}%
\bibitem [{\citenamefont {Badelek}\ \emph {et~al.}(2002)\citenamefont
  {Badelek}, \citenamefont {Kwiecinski},\ and\ \citenamefont
  {Ziaja}}]{Badelek:2002jr}%
  \BibitemOpen
  \bibfield  {author} {\bibinfo {author} {\bibfnamefont {B.~M.}\ \bibnamefont
  {Badelek}}, \bibinfo {author} {\bibfnamefont {J.}~\bibnamefont {Kwiecinski}},
  \ and\ \bibinfo {author} {\bibfnamefont {B.}~\bibnamefont {Ziaja}},\ }\href
  {\doibase 10.1140/epjc/s2002-01041-2} {\bibfield  {journal} {\bibinfo
  {journal} {Eur. Phys. J.}\ }\textbf {\bibinfo {volume} {C26}},\ \bibinfo
  {pages} {45} (\bibinfo {year} {2002})}\BibitemShut {NoStop}%
\bibitem [{\citenamefont {Bass}(2007)}]{Bass:2006dq}%
  \BibitemOpen
  \bibfield  {author} {\bibinfo {author} {\bibfnamefont {S.~D.}\ \bibnamefont
  {Bass}},\ }\href {\doibase 10.1142/S0217732307022967} {\bibfield  {journal}
  {\bibinfo  {journal} {Mod. Phys. Lett.}\ }\textbf {\bibinfo {volume} {A22}},\
  \bibinfo {pages} {1005} (\bibinfo {year} {2007})}\BibitemShut {NoStop}%
\bibitem [{\citenamefont {Breitweg}\ \emph {et~al.}(2000)\citenamefont
  {Breitweg} \emph {et~al.}}]{Breitweg:2000yn}%
  \BibitemOpen
  \bibfield  {author} {\bibinfo {author} {\bibfnamefont {J.}~\bibnamefont
  {Breitweg}} \emph {et~al.} (\bibinfo {collaboration} {ZEUS}),\ }\href
  {\doibase 10.1016/S0370-2693(00)00793-0} {\bibfield  {journal} {\bibinfo
  {journal} {Phys. Lett.}\ }\textbf {\bibinfo {volume} {B487}},\ \bibinfo
  {pages} {53} (\bibinfo {year} {2000})}\BibitemShut {NoStop}%
\bibitem [{\citenamefont {Breitweg}\ \emph {et~al.}(1999)\citenamefont
  {Breitweg} \emph {et~al.}}]{Breitweg:1998dz}%
  \BibitemOpen
  \bibfield  {author} {\bibinfo {author} {\bibfnamefont {J.}~\bibnamefont
  {Breitweg}} \emph {et~al.} (\bibinfo {collaboration} {ZEUS}),\ }\href
  {\doibase 10.1007/s100529901084} {\bibfield  {journal} {\bibinfo  {journal}
  {Eur. Phys. J.}\ }\textbf {\bibinfo {volume} {C7}},\ \bibinfo {pages} {609}
  (\bibinfo {year} {1999})}\BibitemShut {NoStop}%
\bibitem [{\citenamefont {Adloff}\ \emph {et~al.}(1997)\citenamefont {Adloff}
  \emph {et~al.}}]{Adloff:1997mf}%
  \BibitemOpen
  \bibfield  {author} {\bibinfo {author} {\bibfnamefont {C.}~\bibnamefont
  {Adloff}} \emph {et~al.} (\bibinfo {collaboration} {H1}),\ }\href {\doibase
  10.1016/S0550-3213(97)00301-5} {\bibfield  {journal} {\bibinfo  {journal}
  {Nucl. Phys.}\ }\textbf {\bibinfo {volume} {B497}},\ \bibinfo {pages} {3}
  (\bibinfo {year} {1997})}\BibitemShut {NoStop}%
\bibitem [{\citenamefont {Adams}\ \emph {et~al.}(1996)\citenamefont {Adams}
  \emph {et~al.}}]{Adams:1996gu}%
  \BibitemOpen
  \bibfield  {author} {\bibinfo {author} {\bibfnamefont {M.~R.}\ \bibnamefont
  {Adams}} \emph {et~al.} (\bibinfo {collaboration} {E665}),\ }\href {\doibase
  10.1103/PhysRevD.54.3006} {\bibfield  {journal} {\bibinfo  {journal} {Phys.
  Rev.}\ }\textbf {\bibinfo {volume} {D54}},\ \bibinfo {pages} {3006} (\bibinfo
  {year} {1996})}\BibitemShut {NoStop}%
\bibitem [{\citenamefont {Desgrolard}\ \emph {et~al.}(1999)\citenamefont
  {Desgrolard}, \citenamefont {Jenkovszky}, \citenamefont {Lengyel},\ and\
  \citenamefont {Paccanoni}}]{Desgrolard:1999ax}%
  \BibitemOpen
  \bibfield  {author} {\bibinfo {author} {\bibfnamefont {P.}~\bibnamefont
  {Desgrolard}}, \bibinfo {author} {\bibfnamefont {L.~L.}\ \bibnamefont
  {Jenkovszky}}, \bibinfo {author} {\bibfnamefont {A.}~\bibnamefont {Lengyel}},
  \ and\ \bibinfo {author} {\bibfnamefont {F.}~\bibnamefont {Paccanoni}},\
  }\href {\doibase 10.1016/S0370-2693(99)00650-4} {\bibfield  {journal}
  {\bibinfo  {journal} {Phys. Lett.}\ }\textbf {\bibinfo {volume} {B459}},\
  \bibinfo {pages} {265} (\bibinfo {year} {1999})}\BibitemShut {NoStop}%
\bibitem [{\citenamefont {Donnachie}\ and\ \citenamefont
  {Landshoff}(2002)}]{Donnachie:2001zt}%
  \BibitemOpen
  \bibfield  {author} {\bibinfo {author} {\bibfnamefont {A.}~\bibnamefont
  {Donnachie}}\ and\ \bibinfo {author} {\bibfnamefont {P.~V.}\ \bibnamefont
  {Landshoff}},\ }\href {\doibase 10.1016/S0370-2693(02)01556-3} {\bibfield
  {journal} {\bibinfo  {journal} {Phys. Lett.}\ }\textbf {\bibinfo {volume}
  {B533}},\ \bibinfo {pages} {277} (\bibinfo {year} {2002})}\BibitemShut
  {NoStop}%
\bibitem [{\citenamefont {Guler}\ \emph {et~al.}(2015)\citenamefont {Guler}
  \emph {et~al.}}]{Guler:2015hsw}%
  \BibitemOpen
  \bibfield  {author} {\bibinfo {author} {\bibfnamefont {N.}~\bibnamefont
  {Guler}} \emph {et~al.} (\bibinfo {collaboration} {CLAS}),\ }\href {\doibase
  10.1103/PhysRevC.92.055201} {\bibfield  {journal} {\bibinfo  {journal} {Phys.
  Rev.}\ }\textbf {\bibinfo {volume} {C92}},\ \bibinfo {pages} {055201}
  (\bibinfo {year} {2015})}\BibitemShut {NoStop}%
\bibitem [{\citenamefont {Ageev}\ \emph {et~al.}(2007)\citenamefont {Ageev}
  \emph {et~al.}}]{Ageev:2007du}%
  \BibitemOpen
  \bibfield  {author} {\bibinfo {author} {\bibfnamefont {E.~S.}\ \bibnamefont
  {Ageev}} \emph {et~al.} (\bibinfo {collaboration} {Compass}),\ }\href
  {\doibase 10.1016/j.physletb.2007.02.034} {\bibfield  {journal} {\bibinfo
  {journal} {Phys. Lett.}\ }\textbf {\bibinfo {volume} {B647}},\ \bibinfo
  {pages} {330} (\bibinfo {year} {2007})}\BibitemShut {NoStop}%
\bibitem [{\citenamefont {Deshpande}\ \emph {et~al.}(2005)\citenamefont
  {Deshpande}, \citenamefont {Milner}, \citenamefont {Venugopalan},\ and\
  \citenamefont {Vogelsang}}]{Deshpande:2005wd}%
  \BibitemOpen
  \bibfield  {author} {\bibinfo {author} {\bibfnamefont {A.}~\bibnamefont
  {Deshpande}}, \bibinfo {author} {\bibfnamefont {R.}~\bibnamefont {Milner}},
  \bibinfo {author} {\bibfnamefont {R.}~\bibnamefont {Venugopalan}}, \ and\
  \bibinfo {author} {\bibfnamefont {W.}~\bibnamefont {Vogelsang}},\ }\href
  {\doibase 10.1146/annurev.nucl.54.070103.181218} {\bibfield  {journal}
  {\bibinfo  {journal} {Ann. Rev. Nucl. Part. Sci.}\ }\textbf {\bibinfo
  {volume} {55}},\ \bibinfo {pages} {165} (\bibinfo {year} {2005})}\BibitemShut
  {NoStop}%
\bibitem [{\citenamefont {Deshpande}(2017)}]{Deshpande:2017edt}%
  \BibitemOpen
  \bibfield  {author} {\bibinfo {author} {\bibfnamefont {A.}~\bibnamefont
  {Deshpande}},\ }\href {\doibase 10.1142/S0218301317400079} {\bibfield
  {journal} {\bibinfo  {journal} {Int. J. Mod. Phys.}\ }\textbf {\bibinfo
  {volume} {E26}},\ \bibinfo {pages} {1740007} (\bibinfo {year}
  {2017})}\BibitemShut {NoStop}%
\bibitem [{\citenamefont {Helbing}(2006)}]{Helbing:2006zp}%
  \BibitemOpen
  \bibfield  {author} {\bibinfo {author} {\bibfnamefont {K.}~\bibnamefont
  {Helbing}},\ }\href {\doibase 10.1016/j.ppnp.2005.09.003} {\bibfield
  {journal} {\bibinfo  {journal} {Prog. Part. Nucl. Phys.}\ }\textbf {\bibinfo
  {volume} {57}},\ \bibinfo {pages} {405} (\bibinfo {year} {2006})}\BibitemShut
  {NoStop}%
\bibitem [{\citenamefont {Bass}\ and\ \citenamefont
  {Brisudova}(1999)}]{Bass:1997fh}%
  \BibitemOpen
  \bibfield  {author} {\bibinfo {author} {\bibfnamefont {S.~D.}\ \bibnamefont
  {Bass}}\ and\ \bibinfo {author} {\bibfnamefont {M.~M.}\ \bibnamefont
  {Brisudova}},\ }\href {\doibase 10.1007/s100500050228} {\bibfield  {journal}
  {\bibinfo  {journal} {Eur. Phys. J.}\ }\textbf {\bibinfo {volume} {A4}},\
  \bibinfo {pages} {251} (\bibinfo {year} {1999})}\BibitemShut {NoStop}%
\bibitem [{\citenamefont {Bianchi}\ and\ \citenamefont
  {Thomas}(1999)}]{Bianchi:1999qs}%
  \BibitemOpen
  \bibfield  {author} {\bibinfo {author} {\bibfnamefont {N.}~\bibnamefont
  {Bianchi}}\ and\ \bibinfo {author} {\bibfnamefont {E.}~\bibnamefont
  {Thomas}},\ }\href {\doibase 10.1016/S0370-2693(99)00176-8} {\bibfield
  {journal} {\bibinfo  {journal} {Phys. Lett.}\ }\textbf {\bibinfo {volume}
  {B450}},\ \bibinfo {pages} {439} (\bibinfo {year} {1999})}\BibitemShut
  {NoStop}%
\end{thebibliography}%

\end{document}